\documentclass[aps,prb, amsfonts, amssymb, amsmath, superscriptaddress, notitlepage,twocolumn,10pt,nobalancelastpage]{revtex4-2}
\usepackage{graphicx} 
\usepackage{hyperref}
\usepackage{bm}
\hypersetup{ 
    colorlinks=true, 
    linktoc=all,     
    linkcolor=blue,  
    citecolor =blue
} 
\usepackage{amsmath}

\begin{document}

\title{Quantum Criticality of Type-I and Critically Tilted Dirac Semimetals}

\author{Huanzhi Hu}
\affiliation{London Centre for Nanotechnology, University College London, Gordon St., London, WC1H 0AH, United Kingdom}
\affiliation{School of Physics and Astronomy, University of Birmingham, Edgbaston Park Road, Birmingham, B15 2TT, United Kingdom}
\author{Frank Kr\"uger}
\affiliation{London Centre for Nanotechnology, University College London, Gordon St., London, WC1H 0AH, United Kingdom}
\affiliation{ISIS Facility, Rutherford Appleton Laboratory, Chilton, Didcot, Oxfordshire OX11 0QX, United Kingdom}

\begin{abstract}
We investigate the universality of an Ising symmetry breaking phase transition of tilted two-dimensional Dirac fermions, in the type-I phase as well as 
at the Lifshitz transition between a type-I and a type-II semimetal, where the Fermi surface changes from point-like to one with electron and hole pockets
that touch at the overtilted Dirac cones. We compute the Landau damping of long-wavelength order parameter fluctuations 
by tilted Dirac fermions and use the resulting IR propagator as input for a renormalisation-group analysis of the resulting Gross-Neveu-Yukawa
field theory.   We first demonstrate that the criticality of tilted type-I fermions is controlled by a line of fixed points along which 
the poles of the renormalised Green function correspond to an untilted Dirac spectrum with varying anisotropy of Fermi velocities. At the phase 
transition the Lorentz invariance is restored, resulting in the same critical exponents  as for conventional Dirac systems. 
 The multicritical point is given by the endpoint of the fixed-point line. It can be approached along any path in parameter space that avoids the fixed point line 
 of the critical type-I semimetal. We show that the critical exponents at the Lifshitz point are different and that Lorentz invariance is broken. 
 \end{abstract}

\maketitle

\section{Introduction}
\label{sec.intro}

Quantum phase transitions of nodal semimetals with pointlike Fermi surfaces represent the simplest example of fermionic quantum criticality. The symmetry breaking, driven 
by short-ranged repulsive interactions, leads to the opening of a gap in the fermion spectrum and therefore goes hand in hand with a semimetal-to-insulator transition. In the purely relativistic case of 
Dirac fermions it is well understood \cite{Herbut06,Herbut+09,Assad+13,Janssen+14}  that the universal critical behavior is captured by the Gross-Neveu-Yukawa (GNY) theory \cite{Gross+74,ZINNJUSTIN91}  
which describes chiral symmetry breaking and spontaneous mass generation in high-energy physics. 

The coupling between the order parameter fields and the gapless Dirac fermions leads to critical behavior that falls outside the Landau-Ginzburg-Wilson paradigm of a pure order parameter description. 
At the transition, the fermion fields acquire an anomalous dimension, resulting in non-Fermi liquid behaviour, which is the hallmark of fermionic quantum criticality. 

Electron systems can host  quasiparticles at low energies that are more exotic than relativistic Dirac fermions. Examples are semimetals with quadratic band touching points \cite{Fu11,McCann+13,Moon+13,Krempa+14}, 
or semi-Dirac electrons \cite{Pardo+09,Banerjee+09,Kim+15,Kim+17} which display a quadratic band touching along one momentum direction, but behave like relativistic Dirac fermions along
the other direction. Such semi-Dirac electrons occur at the topological phase transition where two Dirac points merge as a result of anisotropy of the tight-binding hopping parameters \cite{Dietl+08,Montambaux+09}.
As one might anticipate, the different low energy electron dispersion in such systems gives rise to novel GNY-type universality classes of symmetry-breaking phase transitions \cite{Roy+18,Sur+19,Uryszek+19,Uryszek+20}.

\begin{figure}[t!]
\centering
\includegraphics[width=0.7\linewidth]{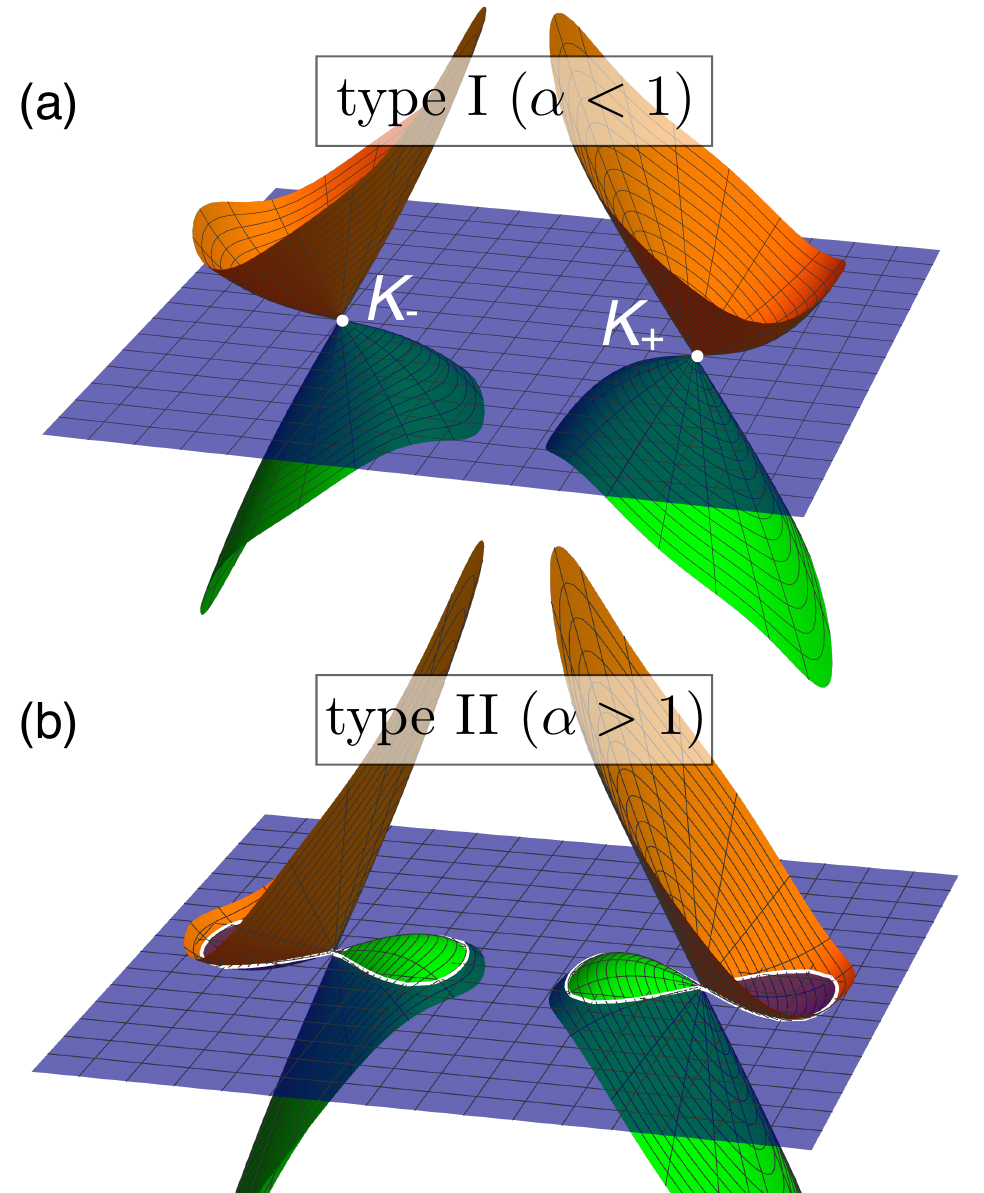}
\caption{Low energy electronic excitations across the type-I to type-II Dirac semimetal transition as a function of the tilt parameter $\alpha$, defined in Eq.~(\ref{eq.model0}). 
(a) In the type-I semimetal the tilt velocity is smaller than the Fermi velocity ($\alpha<1$), resulting in a pair of tilted  Dirac points at valley momenta $\boldsymbol{K}_\pm$. 
(b) In the type-II semimetal the tilt exceeds the Fermi velocity ($\alpha>1$) and the over-tilted Dirac points become touching points of electron and hole Fermi-surface pockets.}
\label{fig1}
\end{figure}

The situation can become even more complicated if the symmetry breaking results in an enlargement of the unit cell on the lattice. In this case, the effective low-energy GNY field theory is enriched by
emergent gauge fields \cite{Christou+20}. Recently, it was pointed out \cite{Seifert+20,Hu+24} that quantum phase transitions of Kitaev quantum-spin liquids, which exhibit fractionalised fermion excitations 
with Dirac or semi-Dirac dispersion, can be understood in terms of similar GNY-type field theories. 

The emergent Lorentz invariance of the Dirac fermions can be broken with more freedom in condensed matter physics. A simple way is to tilt the Dirac spectrum, which introduces anisotropy into the system. 
As the tilt angle increases, the system goes through a Lifshitz transition as the Fermi surface evolves from a single point (type-I Dirac cone) into electron and hole pockets (type-II Dirac cone), as illustrated in 
Fig.~\ref{fig1}. In the type-II phase, the linear dispersions intersect the zero energy axis and form a pair of open electron and hole pockets, which are then cut closed by higher-order terms in the dispersion away 
from the Dirac point. The critical case, which separates the type-I and type-II fermions and which has a flat band at the Fermi surface, is sometimes called a type-III Dirac semimetal. The crucial differences 
of the electronic excitation spectra lead to different quantum transport properties \cite{transport1,transport2,transport3,transport4,transport5}. It is believed that type-II Dirac/Weyl fermions can emerge behind 
the event horizon of a black hole \cite{black_hole,black_hole2,black_hole3}. Tilted Dirac/Weyl semimetals can therefore be used as a platform to simulate the black-hole physics in curved space-time.

While the type-I Dirac/Weyl fermions have already been found in many materials \cite{typeI1,typeI2,typeI3,typeI4}, systems with type-II and type-III Dirac cones remain rare. Nevertheless, it is reported that 
type-II Dirac and Weyl fermions can be found in the family of transition metal dichalcogenides \cite{Huang+16b,Yan+17,Soluyanov+15,Deng+16,Huang+16} and layered oxides \cite{layered_oxide}. 
Type-III Dirac cones are also claimed to emerge in a few materials \cite{typeIII1,typeIII2,typeIII3}. However, the flat bands in such systems are usually sensitive to perturbations. A systematic way to construct 
type-III Dirac cones using a suitable form of perturbations is proposed in Ref.~\cite{typeIIItheory}. An alternative approach is to use photonic lattices as a platform to engineer type-III Dirac cones \cite{photonic1}.
While some materials might be close to the type-I/type-II transition point at ambient conditions, it remains an open experimental challenge to find a way to tune through such a transition. It was proposed that 
the transition might be realised by applying pressure \cite{Rosenstein+23}, tuning the amount of disorder \cite{Park+17}, utilising nitrogen line defects in graphene \cite{Zhang+17}, or by applying external strain 
to two-dimensional nanosheets of phosphorous nitride or AsN \cite{Xie+21}.

The quantum criticality of a tilted type-I Dirac cone has been studied in the context of systems with long-range Coulomb interaction \cite{tilted_coulomb_1,tilted_coulomb_2,tilted_disorder}, symmetry breaking instabilities \cite{tilted_sc,tilted_Ising,tilted_chiral}, and different types of disorder \cite{tilted_disorder,tilted_disorder2}. The consensus is that in the type-I case the tilting parameter is irrelevant under the renormalisation group.
The presence of small electron and hole pockets in the type-II phase significantly complicates the calculation but an RG analysis can nevertheless be carried out by taking into account the proper 
iso-energy contours \cite{tilted_coulomb_1}. In this case the tilting parameter was found to be a relevant perturbation, suggesting that the type-I and type-II semimetals are indeed separated by a critical point. 
It is thus interesting to study multicriticality of this point when sufficiently strong short-range repulsive interactions lead to spontaneous symmetry breaking.

In this paper, we will investigate the nature of symmetry breaking phase transitions of tilted type-I Dirac fermions and of Dirac semimetals that are exactly at the Lifshitz 
transition between the type-I and type-II phases. The remainder of the paper is organised as follows. In Sec.~\ref{sec.model} 
we define a free fermion long-wavelength action that captures the tilt transition between a type-I and a type-II Dirac semimetal, shown in 
Fig.~\ref{fig1}. We then introduce an Ising order parameter field that has a Landau-Ginzburg action and  is linearly coupled to the Dirac fermions via a standard Yukawa interaction. In Sec.~\ref{sec.IR} we consider the 
effect of Landau damping of long-wavelength order parameter fluctuations and determine the asymptotic IR propagator of the order parameter fields. The resulting GNY field theory is analysed in Sec.~\ref{sec.RG},
using a perturbative RG calculation. Finally, in Sec.~\ref{sec.conclusion} we summarise and discuss our results. 

\section{Model}
\label{sec.model}

Our starting point is a zero-temperature free-fermion action that describes the transition between type-I and type-II Dirac fermion semimetals in two spatial dimensions,
\begin{equation}
\mathcal{S}_0  =   \int_{\boldsymbol{k}}  \bar{\Psi}_{\boldsymbol{k}} \left[-i k_0 + \alpha\nu k_x  + (k_x+\epsilon k_x^3)\boldsymbol{\sigma}_x+k_y \boldsymbol{\sigma}_y   \right]\Psi_{\boldsymbol{k}}.
\label{eq.model0}
\end{equation}
Here $k_0$ denotes Matsubara frequency, $k_x$, $k_y$ the spatial momenta, and $\bar{\Psi} = (\bar{\psi}_1,\ldots,\bar{\psi}_N)$ are fermionic Grassmann fields, 
where $\bar{\psi}_n=(\bar{\psi}_{n,A},\bar{\psi}_{n,B})$ are two-component spinors in sub-lattice space.  The $N$ flavours of two component Dirac  fields include spin $s=\uparrow,\downarrow$
and valley index $\nu=\pm 1$. For brevity, we have defined $\boldsymbol{k}=(k_0,k_x,k_y)$ and $\boldsymbol{\sigma}_x$, $\boldsymbol{\sigma}_y$ are Pauli matrices in $A,B$ sublattice space. The resulting electron 
dispersion,
\begin{equation}
\epsilon_{\nu,\pm}(k_x,k_y) = \alpha \nu k_x \pm \sqrt{(k_x+\epsilon k_x^3)^2+k_y^2},
\end{equation} 
contains two parameters, $\alpha$ and $\epsilon$, which correspond to the tilt parameter and curvature correction, respectively. 

The resulting dispersions for the two types of semimetals are plotted 
in Fig.~\ref{fig1}. If the tilt is smaller than the Fermi velocity, $\alpha<1=v_F$, we retain point-like Fermi surfaces, corresponding to a 
type-I Dirac semimetal. In this case the curvature $\epsilon$ can be neglected sufficiently close to the Dirac point. Note that the tilt is of opposite sign for the two valleys, $\nu=\pm 1$.
For $\alpha>1$ the slope along the tilt direction is negative and we are in the type-II phase (see Fig.~\ref{fig1}b).  
As a result of curvature $\epsilon>0$ one obtains closed electron and hole Fermi surface pockets that touch at the overtilted Dirac points.

The inclusion of curvature $\epsilon>0$ seems also essential at the critical tilt $\alpha=1$ since without curvature one would obtain an unphysical flat band at the Fermi level, associated with a divergent density of 
states. Due to curvature the Fermi surface remains point-like at the critical point. However,  because of the different scaling dimension of the cubic $k_x^3$ term the curvature $\epsilon$ will flow to zero 
under RG. As we will see later, we indeed obtain convergent results at the multicritocal point without the inclusion of curvature. 

We couple the Dirac fermions to a dynamical order parameter field through a standard Yukawa coupling, 
\begin{equation}
\mathcal{S}_Y[\phi,\bar{\psi},\psi] = \frac{g}{\sqrt{N}} \int_{\boldsymbol{q},\boldsymbol{k}}  \phi_{\boldsymbol{q}} \bar{\Psi}_{\boldsymbol{k}}
\boldsymbol{\sigma}_z\Psi_{\boldsymbol{k}+\boldsymbol{q}},
\end{equation}
where for simplicity we assume an Ising order parameter, which could for example describe a CDW transition. Note that upon condensation of the order the sublattice symmetry is broken and the fermion 
spectrum becomes gapped. The order parameter field has the standard Landau-Ginzburg $\phi^4$ action, 
\begin{eqnarray}
\label{eq.boson}
\mathcal{S}[\phi] & = &  \frac12 \int_{\boldsymbol{q}} \left( q_0^2 +c_x^2 q_x^2 + c_y^2 q_y^2 +m_\phi^2+\tilde{\Pi}(\boldsymbol{q})  \right) |\phi_{\boldsymbol{q}}|^2\\
& & +\rho \int_{\boldsymbol{q}_1,\boldsymbol{q}_2,\boldsymbol{q}_3}   \phi_{\boldsymbol{q}_1}\phi_{\boldsymbol{q}_2}
\phi_{\boldsymbol{q}_3}\phi_{-\boldsymbol{q}_1-\boldsymbol{q}_2-\boldsymbol{q}_3},\nonumber
\end{eqnarray}
where we have included a boson self-energy correction $\tilde{\Pi}(\boldsymbol{q})$, which is given by the bubble diagram shown in Fig.~\ref{fig2}(a) and will be evaluated in the next section. 

Note that under the RG the fermion self-energy diagram, Fig.~\ref{fig2}(b), generates an additional frequency dependent term of the form $-i \nu k_0\boldsymbol{\sigma}_x$ in the free fermion action
$\mathcal{S}_0[\bar{\psi},\psi]$. We therefore
include such a term in the inverse fermion propagator, 
\begin{equation}
\boldsymbol{G}^{-1}_{\psi,\nu}(\boldsymbol{k}) = -i k_0 + \alpha\nu k_x + (k_x-i\lambda \nu k_0)\boldsymbol{\sigma}_x+k_y \boldsymbol{\sigma}_y,
\end{equation} 
and analyse the coupled RG flow of $\alpha$ and $\lambda$. As one might anticipate, finite $\lambda$ modifies the low-energy electron dispersion, which given by the 
poles of $\boldsymbol{G}_{\psi,\nu}(\boldsymbol{k})$, 
\begin{equation}
\label{eq.disp}
\epsilon_{\nu,\pm}(k_x,k_y) = \nu\frac{\alpha-\lambda}{1-\lambda^2}k_x \pm\sqrt{     \left(  \frac{1-\alpha\lambda}{1-\lambda^2}    \right)^2 k_x^2+\frac{1}{1-\lambda^2}k_y^2}.
\end{equation}
Interestingly, the transition between the type-I and type-II 
semimetals still occurs at the critical value $\alpha=1$ for any value $\lambda<1$.

\section{Landau Damping and IR Propagator}
\label{sec.IR}

The bosonic action $\mathcal{S}[\phi]$ that is generated under perturbative RG is of the conventional Landau-Ginzburg form. However, this neglects the non-analytic bosonic self-energy correction 
$\tilde{\Pi}(\boldsymbol{q}) = \Pi(\boldsymbol{q})-\Pi(\boldsymbol{0})$ due to the Landau damping of the order parameter fluctuations by gapless fermionic particle-hole fluctuations. 
 In $D=2+1$ space-time dimensions the self energy term $\tilde{\Pi}(\boldsymbol{q})$ dominates over the regular quadratic gradient terms in the IR.

While for relativistic Dirac fermions the correct scaling behaviour can be recovered through an order $(1/N)^0$ contribution to the anomalous dimension of the bosonic fields, for systems 
where Lorentz invariance is broken,  it is crucial to use a quadratic bosonic action with inverse propagator $G_{\phi}^{-1}(\boldsymbol{q}) \simeq \tilde{\Pi}(\boldsymbol{q})$
as starting point for subsequent perturbative Wilsonian RG calculation \cite{Isobe+16}.  Using this correct IR scaling form of the propagator, the fluctuation corrections under RG are independent of the 
choice of the UV cut-off scheme and therefore universal \cite{Uryszek+20}.

\begin{figure}[t!]
\centering
\includegraphics[width=0.9\linewidth]{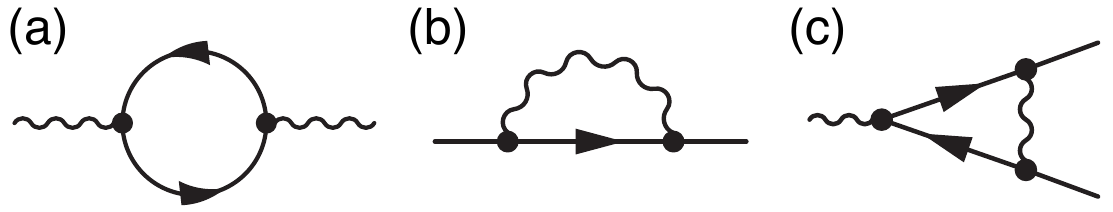}
\caption{(a) Fermionic polarization bubble diagram that gives rise to the non-analytic IR propagator of the bosonic fluctuation field. Panels (b) and (c) show the diagram that contribute to the perturbative renormalisation
 of the free-fermion action and the Yukawa coupling, respectively.}
\label{fig2}
\end{figure}

The bosonic self energy correction [see Fig.~\ref{fig2}(a)] is equal to the integral 
\begin{equation}
\Pi(\boldsymbol{q}) =  \frac{g^2}{N} \int_{\boldsymbol{k}} \textrm{Tr}\Big\{\boldsymbol{G}_\psi(\boldsymbol{k}) \boldsymbol{\sigma}_z \boldsymbol{G}_\psi(\boldsymbol{k}+\boldsymbol{q}) \boldsymbol{\sigma}_z\Big\},
\end{equation}
over frequency and momenta,  where the trace is taken over the $2N$ dimensional fermion flavour space, e.g.  $\textrm{Tr} \left\{ \boldsymbol{\sigma}_i  \boldsymbol{\sigma}_j \right\} = 2N\delta_{ij}$, 
and the fermion Green function for valley $\nu=\pm 1$ is given by 
\begin{equation}
\boldsymbol{G}_{\psi,\nu}(\boldsymbol{k}) = \frac{i(k_0+i\alpha\nu k_x)+(k_x-i\lambda\nu k_0)\boldsymbol{\sigma}_x+k_y\boldsymbol{\sigma}_y}{A(\boldsymbol{k})+2 i (\alpha-\lambda)k_0 k_x},
\end{equation}
with $A(\boldsymbol{k}) = (1-\lambda^2)k_0^2+(1-\alpha^2)k_x^2+k_y^2$. Following the steps of Ref.~\cite{Uryszek+20}, we obtain
\begin{eqnarray}
\tilde{\Pi}(\boldsymbol{q})  & = &   \frac{g^2}{16(1-\alpha\lambda)} \nonumber\\
& & \times \sum_{\nu=\pm 1}\sqrt{(q_0+i\alpha\nu q_x)^2+(q_x-i\lambda\nu q_0)^2+q_y^2},\nonumber
\end{eqnarray}
which after summation over the valleys results in
\begin{eqnarray}
G_{\phi}^{-1}(\boldsymbol{q}) & = &   \frac{g^2}{8\sqrt{2}(1-\alpha\lambda)}  \nonumber\\
& & \times \sqrt{A(\boldsymbol{q})+\sqrt{A^2(\boldsymbol{q}) +4 (\alpha-\lambda)^2 q_0^2 q_x^2}}.
\end{eqnarray}

In the limit $\alpha=\lambda=0$ we recover the known result $G_{\phi}^{-1}(\boldsymbol{q})=\frac{1}{8}g^2 |\boldsymbol{q}|$ for regular Dirac fermions.

\section{Renormalisation Group Analysis}
\label{sec.RG}

We now perform a perturbative RG analysis by integrating out an infinitesimal fraction of UV modes.  Since the universal critical behaviour is independent of the choice of the UV cut-off scheme
we impose a cutoff in frequency-momentum space, 
\begin{equation}
\sqrt{A(\boldsymbol{q})}=\sqrt{(1-\lambda^2)q_0^2+(1-\alpha^2)q_x^2+q_y^2}\le\Lambda,
\end{equation}
and integrate out modes from the infinitesimal shell 
\begin{equation}
\Lambda e^{-d\ell}\le \sqrt{A(\boldsymbol{q})} \le \Lambda.
\end{equation}

 Note that in the type-I semimetal ($\alpha<1$, $\lambda<1$) it is not necessary to 
include an infinitesimal curvature term. We then rescale frequency and momenta as
\begin{equation}
k_0 \to  e^{-z_0 d\ell}k_0, \;\; k_x \to  e^{-z_x d\ell}k_x \;\; \textrm{and}\;\; k_y \to  e^{- d\ell}k_y,
\end{equation}
and fermion and boson fields as
\begin{equation}
\psi \to  e^{-\Delta_\psi/2\, d\ell}\psi, \;\; \phi \to  e^{-\Delta_\phi/2\,d\ell}\phi.
\end{equation}

Let us first consider the renormalisation of the free-fermion action $\mathcal{S}_0[\bar{\psi},\psi]$ due to the fermion self-energy diagram shown in Fig.~\ref{fig2}(b). The corresponding shell 
integral is given by 
\begin{equation}
\boldsymbol{\Sigma}(\boldsymbol{k}) d\ell = -\frac{g^2}{N}  \int_{\boldsymbol{q}}^{>} G_{\phi}(\boldsymbol{q}) \; \boldsymbol{\sigma}_z \boldsymbol{G}_\psi(\boldsymbol{q}+\boldsymbol{k}) \boldsymbol{\sigma}_z.
\end{equation}

Expanding out outer frequency and momenta $\boldsymbol{k}=(k_0,k_x,k_y)$ to linear order we obtain
\begin{eqnarray}
\boldsymbol{\Sigma}(\boldsymbol{k}) & = &  -i k_0 \Sigma_0^{(1)} + \nu k_x \Sigma_x^{(1)} + \left(k_x \Sigma_x^{(2)}-i\nu k_0\Sigma_0^{(2)}\right)  \boldsymbol{\sigma}_x \nonumber\\
& & +k_y \Sigma_y \boldsymbol{\sigma}_y,
\end{eqnarray}
where the fermion self-energy components are given by 
\begin{align}
\Sigma_0^{(1)}d\ell &= \frac{g^2}{N} \int_{\boldsymbol{q}}^{>} G_{\phi}(\boldsymbol{q})\nonumber\\
& \times\left\{\frac{A(\boldsymbol{q})-2\left[(1-\lambda^2)q_0^2 +\alpha(\lambda-\alpha)q_x^2\right]}{A(\boldsymbol{q})^2+4(\alpha-\lambda)^2q_0^2 q_x^2}\right. \nonumber\\
    & \left. + \frac{16(\alpha-\lambda)^2 q_0^2 q_x^2\left[(1-\lambda^2)q_0^2+\alpha(\lambda-\alpha)q_x^2\right]}{\left[A^2(\boldsymbol{q})+4(\alpha-\lambda)^2q_0^2 q_x^2\right]^2} \right. \nonumber\\
    & \left.  - \frac{8(\alpha-\lambda) q_0^2 q_x^2 A(\boldsymbol{q})\left[\alpha-\lambda+\alpha(1-\lambda^2)  \right]}{\left[A^2(\boldsymbol{q})+4(\alpha-\lambda)^2q_0^2 q_x^2\right]^2}   \right\},
\end{align}

\begin{align}
\Sigma_x^{(1)}d\ell &= \frac{g^2}{N} \int_{\boldsymbol{q}}^{>} G_{\phi}(\boldsymbol{q})\nonumber\\
& \times \left\{\frac{\alpha A(\boldsymbol{q})-2\left[(\alpha-\lambda)q_0^2+\alpha(1-\alpha^2)q_x^2\right]}{A(\boldsymbol{q})^2+4(\alpha-\lambda)^2q_0^2 q_x^2}\right. \nonumber\\
    & \left. + \frac{16(\alpha-\lambda)^2 q_0^2 q_x^2\left[(\alpha-\lambda)q_0^2+\alpha(1-\alpha^2)q_x^2\right]}{\left[A^2(\boldsymbol{q})+4(\alpha-\lambda)^2q_0^2 q_x^2\right]^2} \right. \nonumber\\
    & \left.  - \frac{8(\alpha-\lambda) q_0^2 q_x^2 A(\boldsymbol{q})\left[\alpha^2-1+\alpha(\alpha-\lambda)  \right]}{\left[A^2(\boldsymbol{q})+4(\alpha-\lambda)^2q_0^2 q_x^2\right]^2}   \right\},
\end{align}

\begin{align}
\Sigma_x^{(2)}d\ell &= \frac{g^2}{N} \int_{\boldsymbol{q}}^{>} G_{\phi}(\boldsymbol{q}) \nonumber\\
& \times\left\{\frac{A(\boldsymbol{q})-2\left[\lambda(\alpha-\lambda)q_0^2+(1-\alpha^2)q_x^2\right]}{A(\boldsymbol{q})^2+4(\alpha-\lambda)^2q_0^2 q_x^2}\right. \nonumber\\
    & \left. + \frac{16(\alpha-\lambda)^2 q_0^2 q_x^2\left[\lambda(\alpha-\lambda)q_0^2+(1-\alpha^2)q_x^2\right]}{\left[A^2(\boldsymbol{q})+4(\alpha-\lambda)^2q_0^2 q_x^2\right]^2} \right. \nonumber\\
    & \left.  - \frac{8(\alpha-\lambda) q_0^2 q_x^2 A(\boldsymbol{q})\left[(\alpha^2-1)\lambda+\alpha-\lambda \right]}{\left[A^2(\boldsymbol{q})+4(\alpha-\lambda)^2q_0^2 q_x^2\right]^2}   \right\},
\end{align}

\begin{align}
\Sigma_0^{(2)}d\ell &= \frac{g^2}{N} \int_{\boldsymbol{q}}^{>} G_{\phi}(\boldsymbol{q}) \nonumber\\
& \times\left\{\frac{\lambda A(\boldsymbol{q})-2\left[\lambda(1-\lambda^2)q_0^2+(\lambda-\alpha)q_x^2\right]}{A(\boldsymbol{q})^2+4(\alpha-\lambda)^2q_0^2 q_x^2}\right. \nonumber\\
    & \left. + \frac{16(\alpha-\lambda)^2 q_0^2 q_x^2\left[\lambda(1-\lambda^2)q_0^2+(\lambda-\alpha)q_x^2\right]}{\left[A^2(\boldsymbol{q})+4(\alpha-\lambda)^2q_0^2 q_x^2\right]^2} \right. \nonumber\\
    & \left.  - \frac{8(\alpha-\lambda) q_0^2 q_x^2 A(\boldsymbol{q})\left[1-\lambda^2 +\lambda(\alpha-\lambda)  \right]}{\left[A^2(\boldsymbol{q})+4(\alpha-\lambda)^2q_0^2 q_x^2\right]^2}   \right\},
\end{align}
and
\begin{align}
\Sigma_y d\ell &= \frac{g^2}{N} \int_{\boldsymbol{q}}^{>} G_{\phi}(\boldsymbol{q}) \left\{\frac{A(\boldsymbol{q})-2 q_y^2}{A(\boldsymbol{q})^2+4(\alpha-\lambda)^2q_0^2 q_x^2}\right. \nonumber\\
    & \left. \quad\quad\quad+ \frac{16(\alpha-\lambda)^2 q_0^2 q_x^2 q_y^2}{\left[A^2(\boldsymbol{q})+4(\alpha-\lambda)^2q_0^2 q_x^2\right]^2} \right\}.
\end{align}

Keeping the $-i k_0$, $k_x  \boldsymbol{\sigma}_x$ and $k_y  \boldsymbol{\sigma}_y$ terms of the free fermion action scale invariant we obtain
\begin{align}
z_0 &= 1+\Sigma_0^{(1)}-\Sigma_y \\
z_x &=  1+ \Sigma_x^{(2)}-\Sigma_y \\
\Delta_\psi  &= -4-\Sigma_0^{(1)} - \Sigma_x^{(2)} +3 \Sigma_y = -4+\eta_\psi,
\end{align}
where $\eta_\psi$ denotes the anomalous dimension of the fermion fields. With these conventions the terms  $\nu k_x$ 
and $\nu k_0\boldsymbol{\sigma}_x$ won't remain scale invariant for general $\alpha$ and $\lambda$, as described by the resulting 
RG equations 
\begin{align}
\label{eq.alpha}
\frac{d\alpha}{d\ell} &= \Sigma_x^{(1)}-\Sigma_x^{(2)} \alpha,\\
\label{eq.lambda}
\frac{d\lambda}{d\ell}  &= \Sigma_0^{(2)}-\Sigma_0^{(1)}\lambda.
\end{align}

\begin{figure}[t!]
\centering
\includegraphics[width=0.85\linewidth]{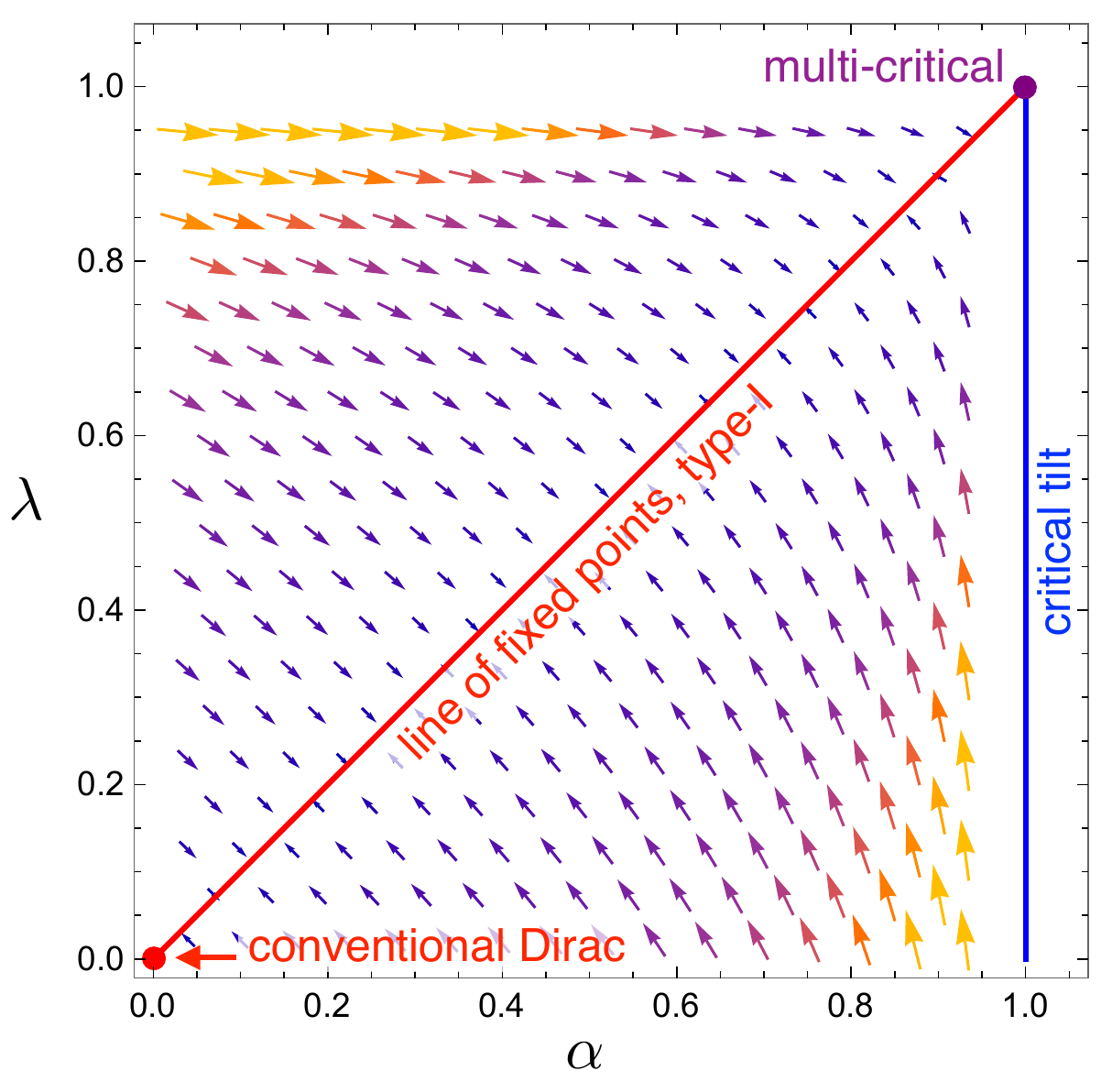}
\caption{RG flow of the parameters $\alpha$ and $\lambda$.}
\label{fig3}
\end{figure}

The renormalisation of the Yukawa coupling $g$ is determined by the diagram shown in Fig.~\ref{fig2}(c). The shell integration results in an 
infinitesimal contribution $dg = g \Omega d\ell$ with 
\begin{align}
\Omega d\ell &= \frac{g^2}{N}  \int_{\boldsymbol{q}}^{>} G_{\phi}(\boldsymbol{q}) \; \boldsymbol{G}_\psi(\boldsymbol{q}) \boldsymbol{\sigma}_z  \boldsymbol{G}_\psi(\boldsymbol{q}) \boldsymbol{\sigma}_z \nonumber\\
&= - \frac{g^2}{N}   \int_{\boldsymbol{q}}^{>} G_{\phi}(\boldsymbol{q}) \;  \frac{A(\boldsymbol{q})}{A(\boldsymbol{q})^2+4(\alpha-\lambda)^2q_0^2 q_x^2}.
\end{align}
Combining with the rescaling we obtain the RG equation
\begin{equation}
\frac{dg}{d\ell} =  \left[-2(1+z_0+z_x) -\Delta_\psi-\frac{\Delta_\phi}{2}+\Omega\right]g.
\end{equation}

Note that it is possible to scale the coupling constant $g$ out of the action by rescaling the bosonic fields, $g\phi\to\phi$. We should therefore impose scale invariance of $g$ which determines the scaling dimension 
of the boson field as 
\begin{equation}
\Delta_\phi = -4-2\left( \Sigma_0^{(1)} +\Sigma_x^{(2)}-\Sigma_y - \Omega \right) = -4 +\eta_\phi. 
\end{equation}

The shell integrals $\Sigma_0^{(1)}$, $\Sigma_0^{(2)}$, $\Sigma_x^{(1)}$, $\Sigma_x^{(2)}$, $\Sigma_y$, and $\Omega$ are non-trivial functions of the tilt parameter $\alpha$ and the emergent 
parameter $\lambda$. In order to identify universal critical behaviour we first need to determine the fixed points of the coupled RG equations (\ref{eq.alpha}) and (\ref{eq.lambda}) for $\alpha$ and $\lambda$.
For initial values $0<\alpha<1$ and $0\le \lambda<1$, corresponding to a type-I Dirac semimetal with tilted Dirac points, the RG flow converges towards a point on the line $\alpha=\lambda<1$, identifying 
it as a line of fixed points.

In the analysis above we have not included the $\phi^4$ vertex $\rho$ of the boson action (\ref{eq.boson}). This is justified since it is irrelevant in the large-$N$ treatment of the GNY fixed point in 2+1 dimensions, 
as can be seen from the scaling dimension $[\rho] = -3(1+z_0+z_x) - 2\Delta_\phi =  -1 +\mathcal{O}(1/N)$.

\subsection{Symmetry-breaking of type-I semimetals}

At first glance it might seem surprising that the symmetry breaking 
phase transition of type-I semimetals is not described by a single fixed point but by a line 
of fixed points. However, it turns out that along the  fixed point line $\alpha=\lambda<1$ the shell integrals that determine the critical exponents remain 
constant and take the analytical values  $\Sigma_0^{(1)}=\Sigma_x^{(2)}=\Sigma_y = 4/(3\pi^2 N)$ and $\Omega = -4/(\pi^2 N)$.
We therefore obtain $z_0=z_x=1$ and 
\begin{equation}
\eta_\psi = \frac{4}{3\pi^2 N},\quad \eta_\phi = -\frac{32}{3\pi^2 N},
\end{equation}
irrespective of the position on this line. This demonstrates that 
the universality of the symmetry-breaking phase transition of tilted type-I semimetals is identical to that of conventional Dirac fermions. Our critical exponents are in agreement with the established results 
for relativistic Dirac fermions in $D=2+1$ dimensions in the large $N$ limit (see Ref.~\cite{Uryszek+20} and references therein). As a note of caution we point out that in our work $N$ refers to the number of Dirac fermion pairs
while in the literature sometimes the total number of fermion flavours, $N_f=2 N$, is used. 

It is also interesting to analyse how the Lorentz invariance in type-I semimetals is restored on large length scales. In fact, from Eq.~\ref{eq.disp} it is apparent that for the 
 free fermion propagator with the additional anomalous $i k_0 \boldsymbol{\sigma}_x$ term,  one can recast the dispersion into the form $\epsilon_{\pm}(k_x,k_y)=\tilde{\alpha} \nu k_x \pm \sqrt{\tilde{v}_x^2 k_x^2+\tilde{v}_y^2 k_y^2}$, where $\tilde{\alpha}=(\alpha-\lambda)/(1-\lambda^2)$ is the effective tilt parameter, and $\tilde{v}_x=(1-\alpha\lambda)/(1-\lambda^2), \tilde{v}_y=1/\sqrt{1-\lambda^2}$ are the effective Fermi velocities. 
 Similar results are also derived in Ref.~\cite{tilted_disorder}. Starting with initially tilted Dirac points, e.g. $0<\alpha<1$ and 
$\lambda=0$, corresponding to the spectrum shown in Fig.~\ref{fig1}(a), the parameters renormalise to values $\alpha_\infty = \lambda_\infty <1$ on the fixed point line, implying $\tilde{\alpha}_\infty=0$. 
We see that though $\alpha$ flows to some non-zero fixed point, the effective tilt parameter $\tilde{\alpha}$ is irrelevant under RG. 
On the fixed point line the renormalised electron dispersion takes the form $\epsilon(k_x,k_y)=\pm \sqrt{\tilde{v}_x^2 k_x^2+\tilde{v}_y^2 k_y^2}$, which describes untilted Dirac cones but with anisotropic Fermi velocities. Such an anisotropy can be simply scaled out by a rescaling of length along one of the coordinate axis. Hence the emergent Lorentz invariance is restored.

We briefly discuss how our results are related to relevant previous work. 
In Ref.~\cite{tilted_Ising} the authors studied the criticality of the type-I tilted Dirac cone with the same Ising Yukawa coupling through a $(4-\epsilon)$ expansion, and found that the tilt parameter $\alpha$ is marginal. 
 The difference in the renormalisation of $\alpha$ is due to the different forms of the order parameter propagator $G_{\phi}$. 
 In the $\epsilon$ expansion, the Landau damping of $G_{\phi}$ is of order $\sim \boldsymbol{k}^2$, indicating its perturbative nature, and hence one should keep the original bosonic propagator and the 
$\phi^4$ term in the RG calculation. In this case the tilting term indeed commutes with the rest of the action and cannot be renormalised by the Yukawa coupling. 
In $(2+1)$ dimensions, on the other hand, the Landau damping leads to a non-perturbative form of the bosonic self-energy $\sim |\boldsymbol{k}|$, 
rendering the bare ($\sim \boldsymbol{k}^2$) propagator sub-leading and the $\phi^4$ irrelevant. 
With the damped bosonic propagator, the Yukawa coupling will renormalise the effective tilt parameter $\tilde{\alpha}$ and render it irrelevant. The irrelevance of the tilt in the type-I case is also found in Refs.~\cite{tilted_coulomb_1,tilted_coulomb_2}, where the bosonic propagator in $(2+1)$ dimensions takes a similar form $\sim |\boldsymbol{k}|$ due to its origin in the long-range Coulomb interaction.

\subsection{Multicriticality}

We can evaluate the universal critical behaviour of symmetry breaking at the type-I/type-II transition by approaching the point $\alpha=\lambda=1$ 
on any path that does not follow the fixed point-line of the type-I semimetal, e.g. we could consider $\alpha=1-\delta$ and $\lambda = (1-\delta)\alpha$
in the limit $\delta\to 0$. 

Even without the inclusion of infinitesimal curvature the shell integrals converge to $\Sigma_0^{(1,2)}  \approx 0.1644/N$, $\Sigma_x^{(1,2)}  \approx 0.1238/N$, $\Sigma_y \approx 0.1401/N$, 
and $\Omega \approx -0.4149/N$, which results in the critical exponents
\begin{align}
z_0 &= 1+\frac{0.0242}{N}, \quad\eta_\psi = \frac{0.1322}{N},\\
z_x &= 1-\frac{0.0163}{N},  \quad\eta_\phi = -\frac{1.1259}{N}.
\end{align}

Similar to the type-I case, the non-zero anomalous dimension of the fermion fields suggests non-Fermi liquid behaviour at the multicritical point, where the quasiparticle residue scales to zero at low energy limit as a power law of the energy scale. At the multicritical point, the values of the critical exponents are slightly different from the type-I case, which indicates different scaling behaviours of physical observables in comparison to isotropic 
Dirac fermions. The deviation of the scaling exponents $z_0$ and $z_x$ from one indicates the breaking of Lorentz invariance.

\section{Conclusions}
\label{sec.conclusion}

In this paper we have studied the nature of symmetry-breaking phase transitions 
of two-dimensional tilted Dirac fermions, in the type-I phase and  
exactly at the 
Lifshitz transition between a type-I and type-II semimetal, sometimes referred to as type-III
fermions. At the type-I/type-II transition the Fermi surface changes from point-like to one 
composed of small electron and hole pockets that are attached to the overtilted Dirac cones.  
  
For simplicity, we focussed on an Ising order parameter which could for example 
describe a CDW transition where the charge on the two sublattices becomes unequal.  
Such a transition could be driven by sufficiently strong repulsions between 
fermions on neighbouring lattice sites.  The generalisation to different order parameters, 
such as superconductivity or antiferromagnetism, is straightforward since the main 
computational challenges stem from the nature of the fermionic excitations at the type-I/type-II
transition, which do not depend on the number of bosonic order-parameter components. 

We treated the problem using a renormalisation-group (RG) analysis of a  Gross-Neveu-Yukawa 
type field theory which describes the coupling between the gapless fermion excitations of the tilted Dirac 
semimetal to the bosonic order parameter fluctuations through a standard Yukawa coupling. This coupling 
gives rise to Landau damping of bosonic order parameter fluctuations by electronic particle-hole fluctuations. 
It is crucial to include the resulting bosonic self-energy correction since it dominates 
over the conventional gradient terms in the IR limit in 2+1 space-time dimensions and therefore 
determines the correct scaling form of the bosonic propagator. 

The Landau damping, which was neglected in previous investigations of tilted 
Dirac semimetals, leads to an important mixing between the Dirac valleys in the Brillouin 
zone. In systems with inversion symmetry tilted Dirac cones always occur in pairs, 
at opposite momenta and with opposite tilt. Both of these valleys contribute equally to the Landau 
damping of the bosonic order-parameter fluctuations which in turn contribute to the renormalisation 
of the fermion propagator at each valley. 

We found that in the type-I phase the tilt parameter does not simply flow to zero under the RG. Instead 
an additional linear-frequency term is generated in the free fermion action. This process can be viewed as 
a gradual absorption of tilt by rotating between Matsubara frequency and the spatial momentum plane. We 
identified a line of fixed points along which the poles of the renormalised Green function correspond to an 
untilted Dirac fermion spectrum with changing anisotropy of Fermi velocities. Such an anisotropy  does not
change the nature of the phase transition since it can be simply scaled out. This result demonstrates  
that Lorentz invariance is restored on large length and time scales and that the universality of 
symmetry-breaking phase transitions of tilted type-I Dirac semimetals is the same as that of 
conventional untilted Dirac systems. 

The multicritical point corresponds to the endpoint of the line of fixed points at the critical tilt value.  
We were able to obtain the critical exponents of the Ising transition at the Lifshitz point  
by approaching the multicritical point along any chosen path that does not follow the fixed-point line 
of the type-I semimetal. Our results show that Lorentz invariance is broken at the multicritical point. 

{\bf Acknowledgements:} We thank Jacqueline Bloch and Mike Gunn for fruitful discussions.

\end{document}